\documentclass[manuscript]{aastex}

\newcommand{\myemail}{k-heike@astro.phys.sci.ehime-u.ac.jp}

\slugcomment{}

\shorttitle{Variable X-ray sources in NGC 1569}
\shortauthors{Heike et al.}

\begin{document}


\title{Discovery of Bright Variable X-ray Sources in NGC 1569 with $Chandra$}


\author{K. Heike, H. Awaki, Y. Misao}
\affil{Department of Physics, Faculty of Science, Ehime University,
    Matsuyama, 790-8577, Japan}
\email{\myemail}

\author{K. Hayashida}
\affil{Department of Astrophysics, Faculty of Science, Osaka University, Toyonaka, 560-0043,
Japan}

\and

\author{K. A. Weaver}
\affil{NASA Goddard Space Flight Center, Code 662, Greenbelt, MD 20771}




\begin{abstract}

From the analysis of a $\sim$100 ks $Chandra$ observation of the
dwarf starburst galaxy NGC 1569,
we have found that the X-ray point sources, CXOU 043048.1+645050 and
CXOU 043048.6+645058, showed significant time variability.  During
this observation, the X-ray flux of CXOU 043048.1+645050 increased by
10 times in only 2 $\times$ 10$^{4}$ s.  Since the spectrum in its bright
phase was fitted with a disk blackbody model with $kT_{\rm in}\approx$ 0.43 keV 
and the bolometric luminosity is $L_{\rm bol}\approx$ 10$^{38}$ ergs s$^{-1}$, 
this source is an X-ray binary with a stellar mass black-hole.  
Since the spectrum in its faint phase was also fitted
with a disk blackbody model, the time variability can be explained by a 
change of the accretion rate onto the black hole. The other variable source, 
CXOU 043048.6+645058, had a flat spectrum with a photon index of
$\sim$1.6. This source may be an X-ray binary with an X-ray luminosity of
several $\times$ 10$^{37}$ ergs s$^{-1}$.
In addition, three other weak sources showed possible
time variability.  Taking all of the variability into account
may suggest an abundant population of compact X-ray sources in NGC 1569.

\end{abstract}




\section{Introduction}
Starburst galaxies emit hard X-rays as 
wells as soft X-rays from the hot gas due to starburst phenomena (e.g. Pietsch et al. 2000, Bregman, Schulman, \& Tomisaka 1995). 
The hard X-rays can be represented 
by thermal emission with $kT$ $>$ a few keV (e.g. Awaki et al. 1996, Tsuru et al. 1997). Although the $ASCA$ observations
did not have the capability to investigate the hard X-rays in detail, the 
$Chandra$ X-ray observatory, 
with its superior spatial resolution, has now resolved the hard X-ray 
emission into many discrete sources (e.g. Griffiths et al. 2000 for M82 
and Weaver et al. 2001 for NGC 253).
Recent $Chandra$ observations suggest the existence of a relationship 
between star formation and these discrete sources, especially 
ultra-luminous X-ray sources (e.g. Fabbiano, Zezas, \& Murray 2001; Lira et al. 2002). 
The relationship is crucial to study the evolution of starburst galaxies. 
Thus, we have to know the nature of discrete sources in 
detail.

NGC 1569 is one of the closest dwarf starburst galaxy at a distance of 
2.2 Mpc with similar metallicity to that of the Magellanic clouds 
(Israel 1988). The size of this galaxy is 1.6 $\times$ 1.1 kpc. 
Since this galaxy is thought to have entered a starburst phase
10--20 Myr ago (e.g. Israel \& De Bruyn 1988), it is suitable for the 
investigation of  
the relationship between starburst phase and the X-ray populations.  
The ASCA observation revealed that NGC 1569 possesses a hard X-ray component with 
$kT\approx$ 3.7 keV similar to
other starburst galaxies (Della Ceca et al. 1996). The hard component
was spatially resolved into point sources by the $Chandra$ observation
(Martin, Kobulnicky, \& Heckman 2002): in order to properly classify
these sources, we have searched for short timescale variability of
the hard X-ray sources in NGC 1569. Here, we present our results and
discuss their implication for the X-ray sources in NGC 1569.

\section{Observations and Results}
NGC 1569 was observed with the $Chandra$ X-ray observatory on 2000 April
11. The galaxy was placed on the ACIS-S3 chip. The data were screened,
and were available in the archive under sequence number of 600085.
The total exposure time with all known corrections was $\sim$97 ks.

We analyzed the data of bright X-ray sources in the central region
of NGC 1569 using CIAO 2.2.1. An X-ray image in the 0.5--10 keV band of
the NGC 1569 central region was produced, and then the image was smoothed
using the adaptive smoothing algorithm in CSMOOTH. The smoothing scale
was adjusted to achieve a minimum S/N of 2 and a maximum 
S/N of 5. The image of the  central 
region of NGC 1569 is shown in Fig.1. The image size is a
31.$^{\prime\prime}$4 square, which corresponds to a 335 pc square at
NGC 1569. Four bright sources, 
(CXOU 043048.1+645050, CXOU 043048.2+645046, CXOU 043048.6+645058, and
CXOU 043049.8+645055), with S/N $>$ 5, are clearly detected. 

We extracted light curves of all four sources with a bin size of 2000 s
(Fig.2). The data were accumulated within 4$^{\prime\prime}$ diameter
circles centered on the sources. Within this diameter, the encircled energy is
greater than 90\%.  We found that the X-ray flux of CXOU 043048.1+645050
increased by $\sim$10 times in only 2 $\times$ 10$^{4}$ s, and
that CXOU 043048.6+645058 showed rapid variability with a time scale of
2000 s, and a flare-like event towards the end of the observation in Fig. 2.
In order to examine these time variabilities, we made a
corresponding light curve of a nearby
background region (see Fig.1). The count rate of the background was scaled
by the size of the source region for comparison with the source count rate.
The averaged background count rate was only 4 $\times$ 10$^{-4}$ c s$^{-1}$,
which was much smaller than the source count rates.

We divided the observation into two, and then extracted spectra of the two
variable sources in order to investigate whether any spectral changes
accompany the time
variability. The first half of the observation (with an accumulation time
of $\sim$48 ks) corresponds to a weak phase for both of the sources.
The surface brightness of the diffuse X-ray emission from
the galaxy is very low (Fig. 1).  Therefore a background spectrum
(shown in Fig. 2) was extracted
from the data within the nearby blank-field region displayed in Fig.1 to
subtract the diffuse X-ray emission.
We note that the background was high at the start of the observation
as pointed out by Martin, Kobulnicky, \& Heckman (2002). 
However, we included these events
in our analysis, because the background count rate was still much smaller
than the source count rate.
Fig. 3 shows spectra of CXOU 043048.1+645050 and CXOU 043048.6+645058
from the first half and in the latter half of the observation.
The spectra were fitted  separately with a single power-law model. 
In the fitting, the column density was fixed at Galactic column density 
of $N_{\rm H}$ = 2.1 $\times$ 10$^{21}$ cm$^{-2}$ (Burton 1985), 
when the best-fit column was smaller than the Galactic value. 
The observed X-ray fluxes and absorbed luminosities are
estimated using the best-fit values (Table 1).

\section{Discussion}

In the galaxy NGC 1569, four bright hard X-ray sources are detected with 
$Chandra$.  Two of them show short-timescale variability. Their mean 
luminosities are of order 10$^{37}$ ergs s$^{-1}$ in the 0.5--10 keV band, 
and their spectra are well fitted with a single power-law model.

CXOU 043048.1+645050 has no optical counterpart, and was considered to
be an X-ray binary by Martin, Kobulnicky, \& Heckman (2002) 
from its spectral energy density.
Its X-ray flux dramatically increased in only 2 $\times$ 10$^{4}$ s. This 
short timescale variability suggests that this source is actually associated 
with an X-ray binary, although it has a steep spectrum with $\Gamma \approx$ 4.1. 
The mean, absorbed luminosity 
in the latter half of the $Chandra$ observation was 5.3 $\times$ 10$^{37}$
ergs s$^{-1}$ in the 0.5--10 keV band (Table 1).  In its bright phase,
which occurred 8 $\times$ 10$^{4}$ s after the begining of 
the observation (in Fig. 2), the absorbed 0.5--10 keV luminosity
was $\sim$1 $\times$ 10$^{38}$ ergs s$^{-1}$, which translates to
an unabsorbed luminosity of 5.4 $\times$ 10$^{38}$
ergs s$^{-1}$.  This is larger than the Eddington limit for a 1.4 $M_{\sun}$
neutron star, $L_{\rm {E}}^{\rm {NS}}$ = 2 $\times$ 10$^{38}$ ergs s$^{-1}$
(Supper et al. 1997). Most of high mass X-ray binaries have flat X-ray
spectra with a photon index of $\sim$1 (e.g. Yokogawa et al. 2000) and 
most of low mass X-ray binaries (hereafter LMXBs) have an age of more than 
100 Myr (e.g. White \& Ghosh 1998). 
Therefore, this source seems likely to be a 
stellar black hole rather than a neutron star binary. 
We fitted the spectrum
with a disk blackbody model, and then obtained the best-fit values of
$kT_{\rm in}$ = 0.43$\pm$0.05 keV and $N_{\rm H}$ = (2.1$\pm$0.4) $\times$ 10$^{21}$ 
cm$^{-2}$, where $T_{\rm in}$ is the temperature at the innermost 
disk boundary (Table 2).
The bolometric luminosity of the disk blackbody component 
($L_{\rm bol}$) in the bright phase was estimated
to be 1 $\times$ 10$^{38}$(cos $i$)$^{-1}$ ergs s$^{-1}$, 
where $i$ is the inclination angle of the disk (Makishima et al. 2000). 
The $T_{\rm in}$ and $L_{\rm bol}$ suggest that the source is a possible 
candidate of a stellar black hole with $\sim$20(cos $i$)$^{-\frac{1}{2}}$ 
$M_{\odot}$. The spectrum in the first half was also fitted
with a disk blackbody model with $kT_{\rm in} = 0.27^{+0.13}_{-0.08}$ keV 
and $L_{\rm bol}\approx$ 1.3 $\times$ 10$^{37}$(cos $i$)$^{-1}$ ergs s$^{-1}$. These values can be 
explained by a smaller mass-accretion onto the stellar black hole.

The large values of $\chi^{2}$ in Tables 1 and 2 are mostly due to 
absorption-like 
features at about 1.2 keV and 1.6 keV. These features exist before background 
subtraction and so are not caused by problems with the background spectra.
Adding absorption lines at 1.2 keV and 1.6 keV to the power law and disk 
blackbody models improves the reduced $\chi^{2}$ values in tables 1 and 2 to 
1.1 and 1.2, respectively. Future observations to confirm whether these 
absorption features are intrinsic or not are very important.

CXOU 043048.6+645058 is associated with a star cluster (Hunter et al. 2000).
The age of the star cluster is considered to be $<$ 30 Myr, based on
the optical color of the cluster. CXOU 043048.6+645058 showed variability on the order of 
a thousand seconds, and also showed an interesting flare-like feature. 
However, its spectrum did
not change significantly during the $Chandra$ observation.
The unabsorbed luminosities in the 0.5--10 keV band were 3.6 $\times$ 10$^{37}$ 
ergs s$^{-1}$
and 7 $\times$ 10$^{37}$ ergs s$^{-1}$ in the first and latter half of the
observation, respectively.
Although the X-ray spectrum and the X-ray luminosity are similar to those of
neutron star binaries, the presented evidence does not allow to distinguish 
between the two possibilities of a neutron star binary and a black-hole binary.

We found that several weak sources also showed time variability. Fig. 4
shows the X-ray images during the first half and the latter half of the
observation. Sources \#1, \#2, and \#3 are seen in the first half, but
are not seen in the latter half. Their count rates during the first half
were (2$\pm$1) $\times$ 10$^{-4}$ c s$^{-1}$,
(2$\pm$1) $\times$ 10$^{-4}$ c s$^{-1}$, and 
(7$\pm$2) $\times$ 10$^{-4}$ c s$^{-1}$, which correspond to the
X-ray luminosities in the range of (1--4) $\times$ 10$^{36}$ ergs s$^{-1}$, 
and those during the latter half were 
$<$ 1 $\times$ 10$^{-4}$ c s$^{-1}$, $<$ 1 $\times$ 10$^{-4}$ c s$^{-1}$, 
(2$\pm$1) $\times$ 10$^{-4}$ c s$^{-1}$, respectively. 
These errors are given for 68\% confidence.
The time variabilities suggest that  these sources are X-ray binaries. 
Since many LMXBs are formed after few Gyr from the start of 
starburst phase, most of those detected sources would be high mass 
X-ray binaries rather than LMXBs. In fact, the similar irregular galaxy, 
SMC contains many high mass X-ray binaries (Yokogawa et al. 2000).

\section{Conclusion}
We have discovered rapid variability from two bright X-ray
point sources in NGC 1569 with
$Chandra$. The X-ray flux of CXOU 043048.1+645050 increased within
2 $\times$ 10$^{4}$ s. During its bright phase, the spectrum was fitted 
with a disk blackbody model
with $kT_{\rm in}\approx$ 0.43 keV and the bolometric luminosity is $\sim$10$^{38}$ ergs s$^{-1}$.
Based on its properties, we suggest that this object may be a stellar 
black hole binary rather than a neutron star binary.
CXOU 043048.6+645058 also showed short time variability along
with a flare-like event. Its spectrum did not significantly change
during this time. Considering the X-ray characteristics of this
source, it may be a neutron star binary, although we can not rule out
a possibility of a stellar black hole binary.

We also found that three weak X-ray sources showed possible time variability.
This large number of variable sources
may suggest an abundant population of X-ray binaries in NGC 1569.
Using a large X-ray observatory, such as $Constellation$-$X$ and $XEUS$, we
can will be able to detect the time variability of fainter objects, and may
detect X-ray pulsations of X-ray binaries in NGC 1569.

\acknowledgments
The authors wish to thank for the staff of $Chandra$ data center.
We also thank an anonymous referee for comments and suggestions that greatly 
improved the paper.
This study was carried
out as a part of the ``Ground Research Announcement for Space Utilization''
promoted by the Japan Space Forum (HA) and in part by the
Grant-in-Aid for Scientific Research (14340061) of the Ministry of Education,
Culture, Sports, Science and Technology (KH \& HA).



\clearpage


\begin{figure}
\plotone{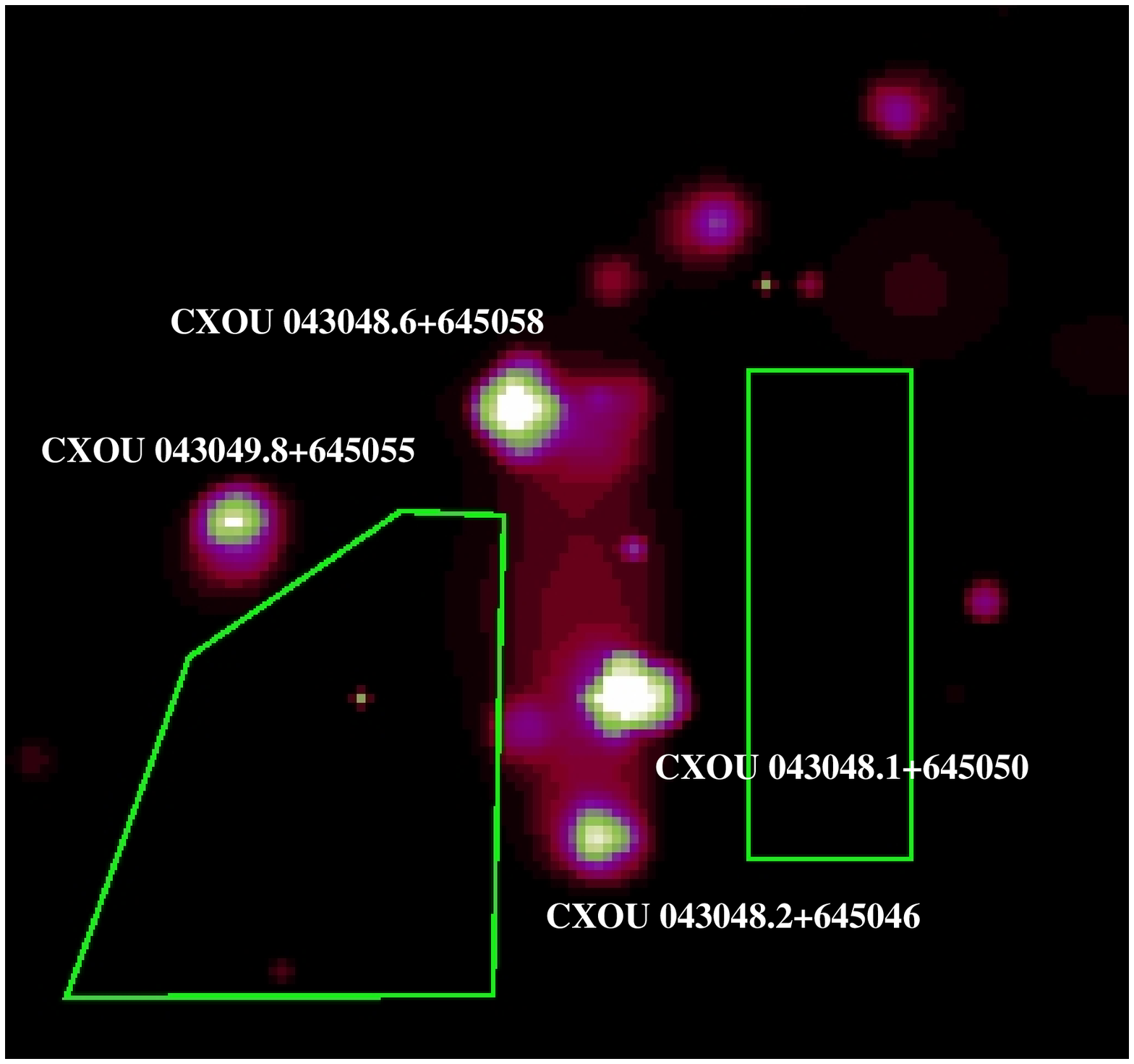}
\caption{A $Chandra$ X-ray image of the central region (31.$^{\prime\prime}$4
$\times$ 31.$^{\prime\prime}$4) in NGC 1569. The image was smoothed using the
adaptive smoothing algorithm in CSMOOTH. The smoothing scale was adjusted
to achieve a minimum S/N of 2 and a maximum S/N of 5. 
We also indicate the background region for timing and spectral analysis
by the green solid lines.
\label{fig1}}
\end{figure}

\clearpage

\begin{figure}

 \begin{center}

  \includegraphics[width=5cm,keepaspectratio,clip]{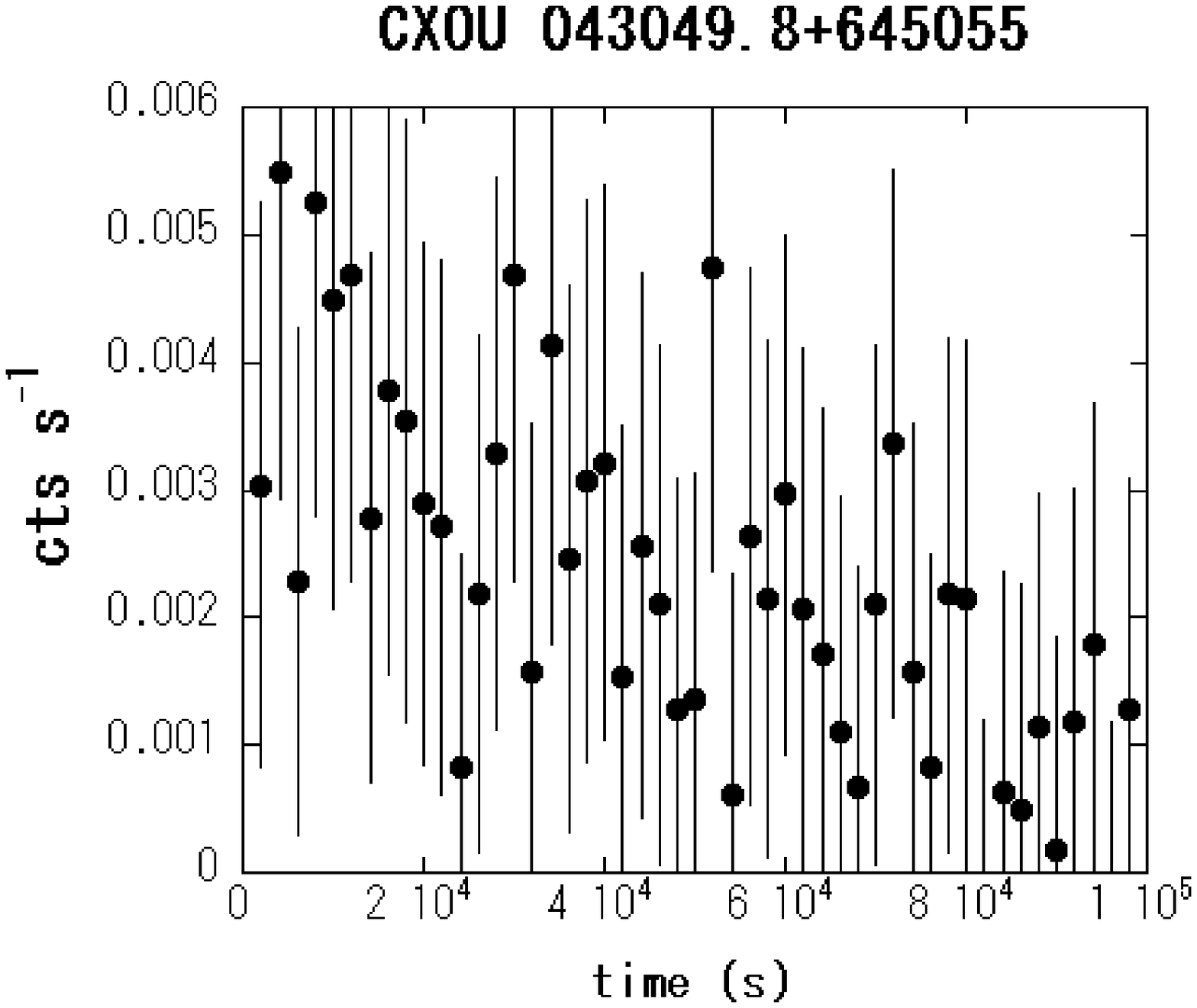}
  \includegraphics[width=5cm,keepaspectratio,clip]{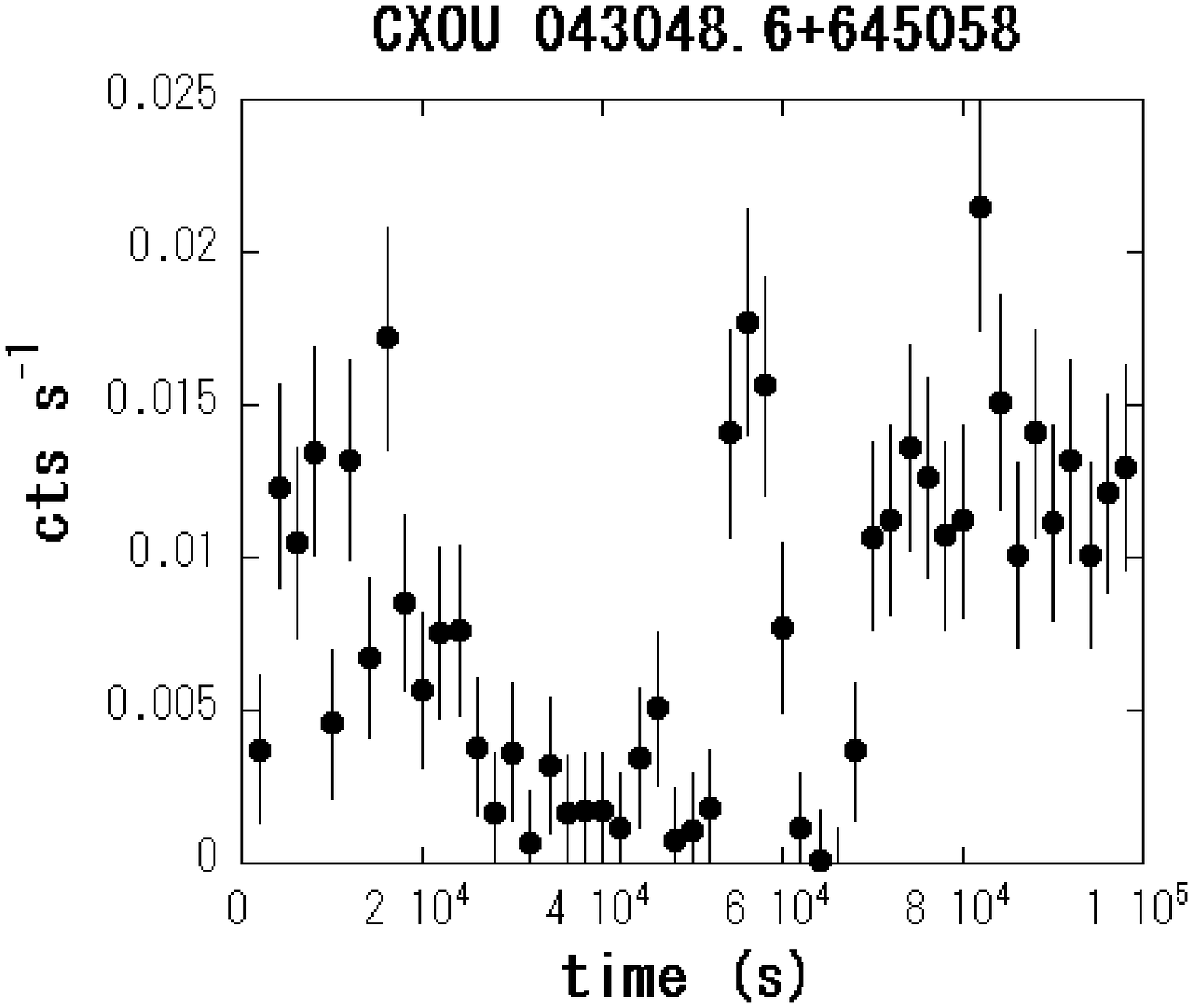}
  \includegraphics[width=5cm,keepaspectratio,clip]{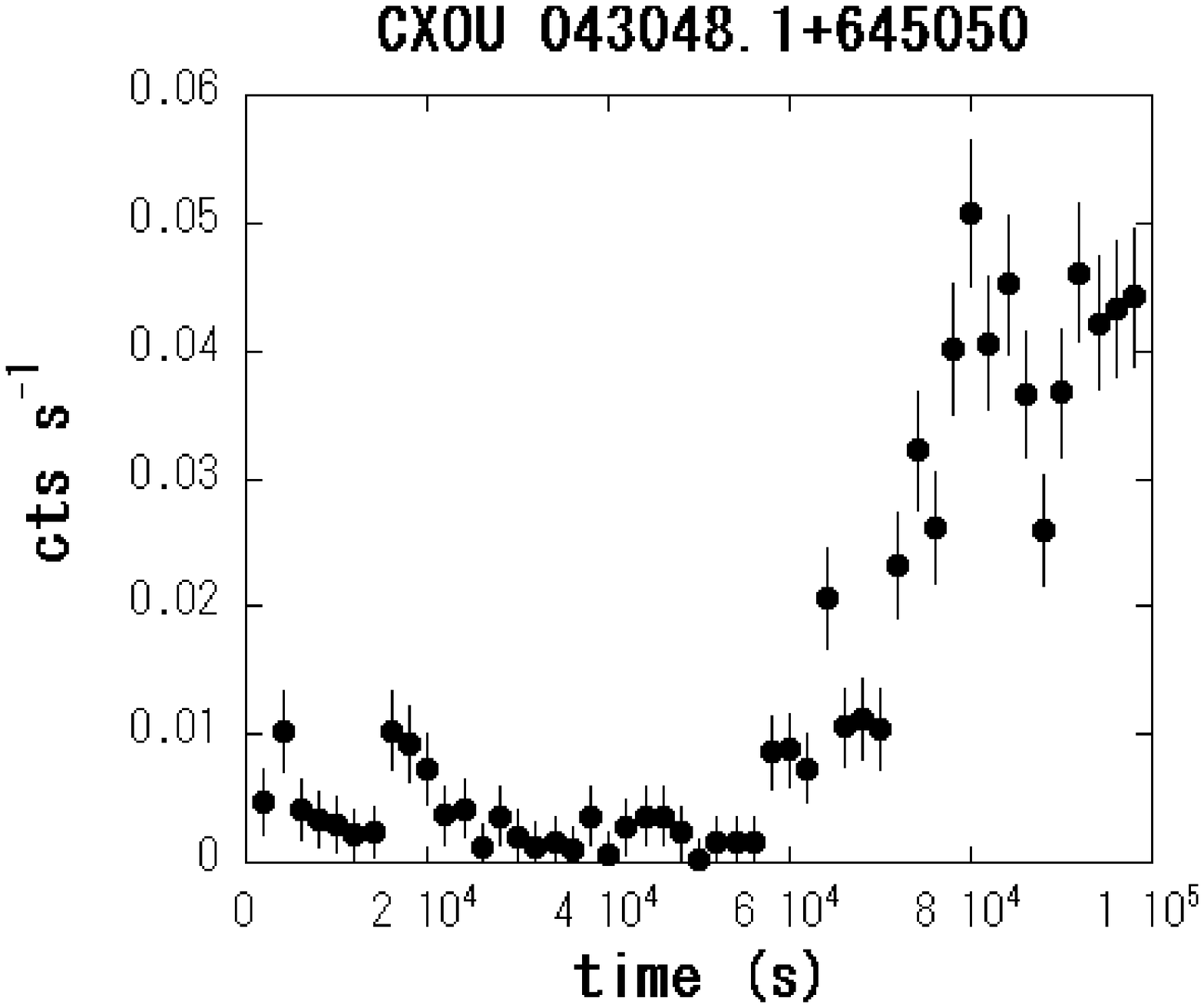}
  \includegraphics[width=5cm,keepaspectratio,clip]{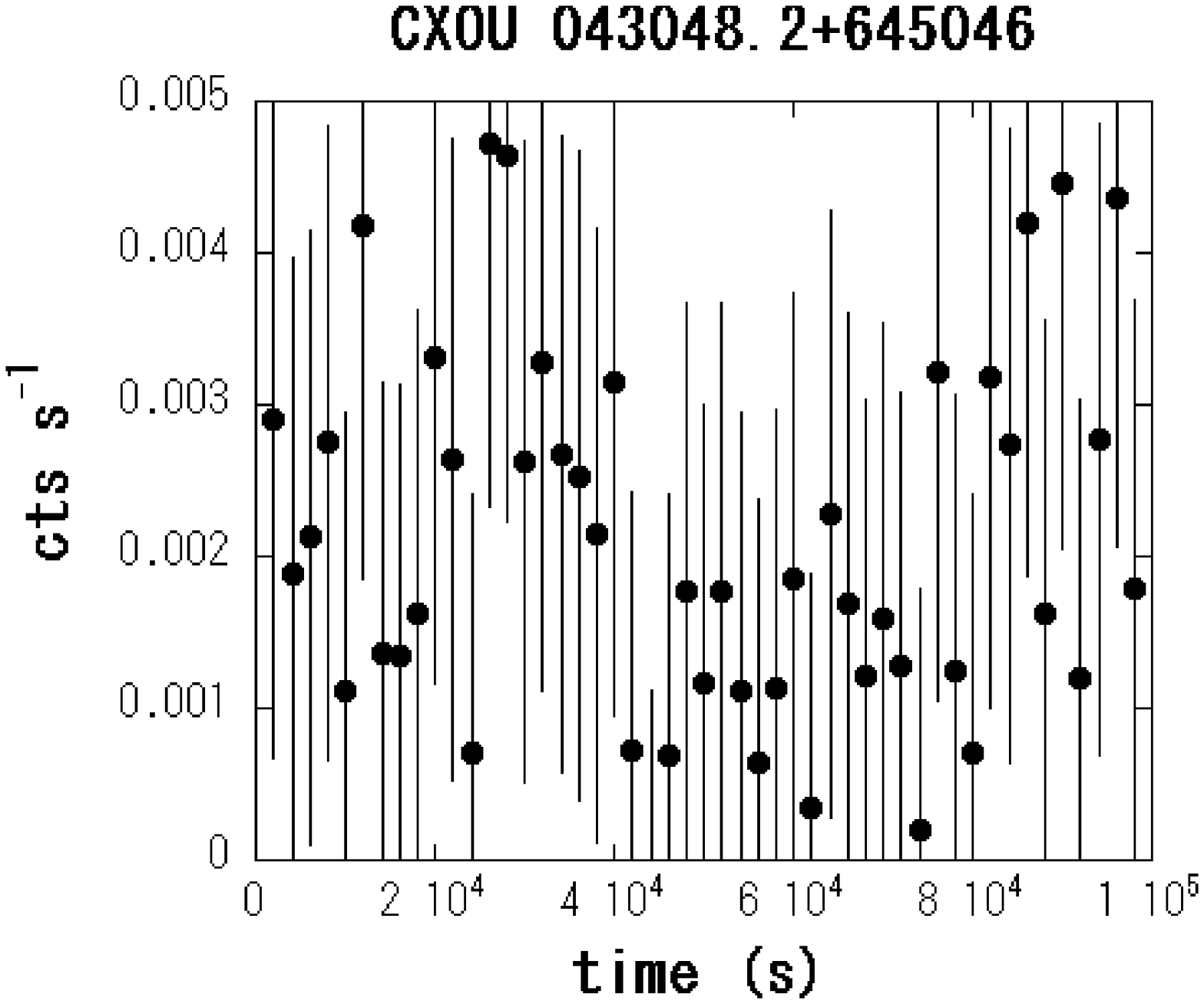}
  \includegraphics[width=5cm,keepaspectratio,clip]{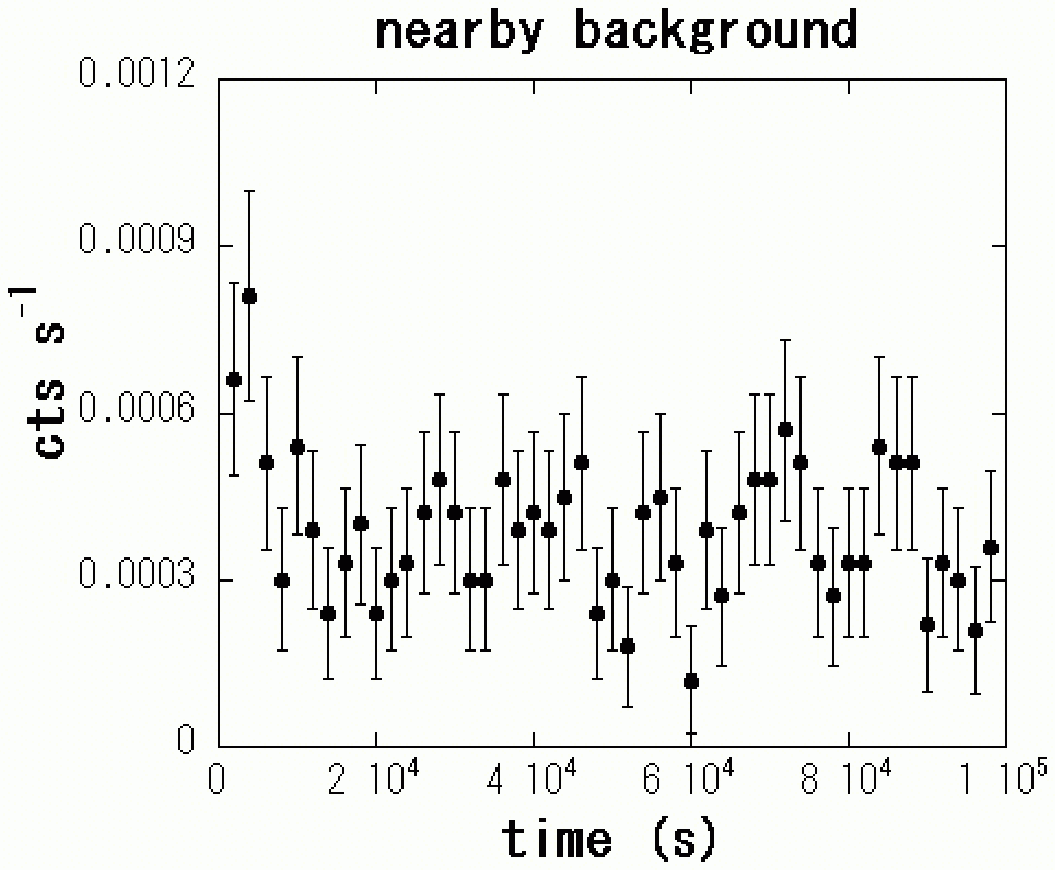}
 \end{center}
\caption{The light curves of the four bright X-ray sources and the background
in the 0.5 -- 10 keV band. The observation was started at 00:05:07(UT) on
2000 April 11. The time bin was 2000 s.
\label{fig2}}
\end{figure}

\clearpage

\begin{figure}
\begin{center}
\plottwo{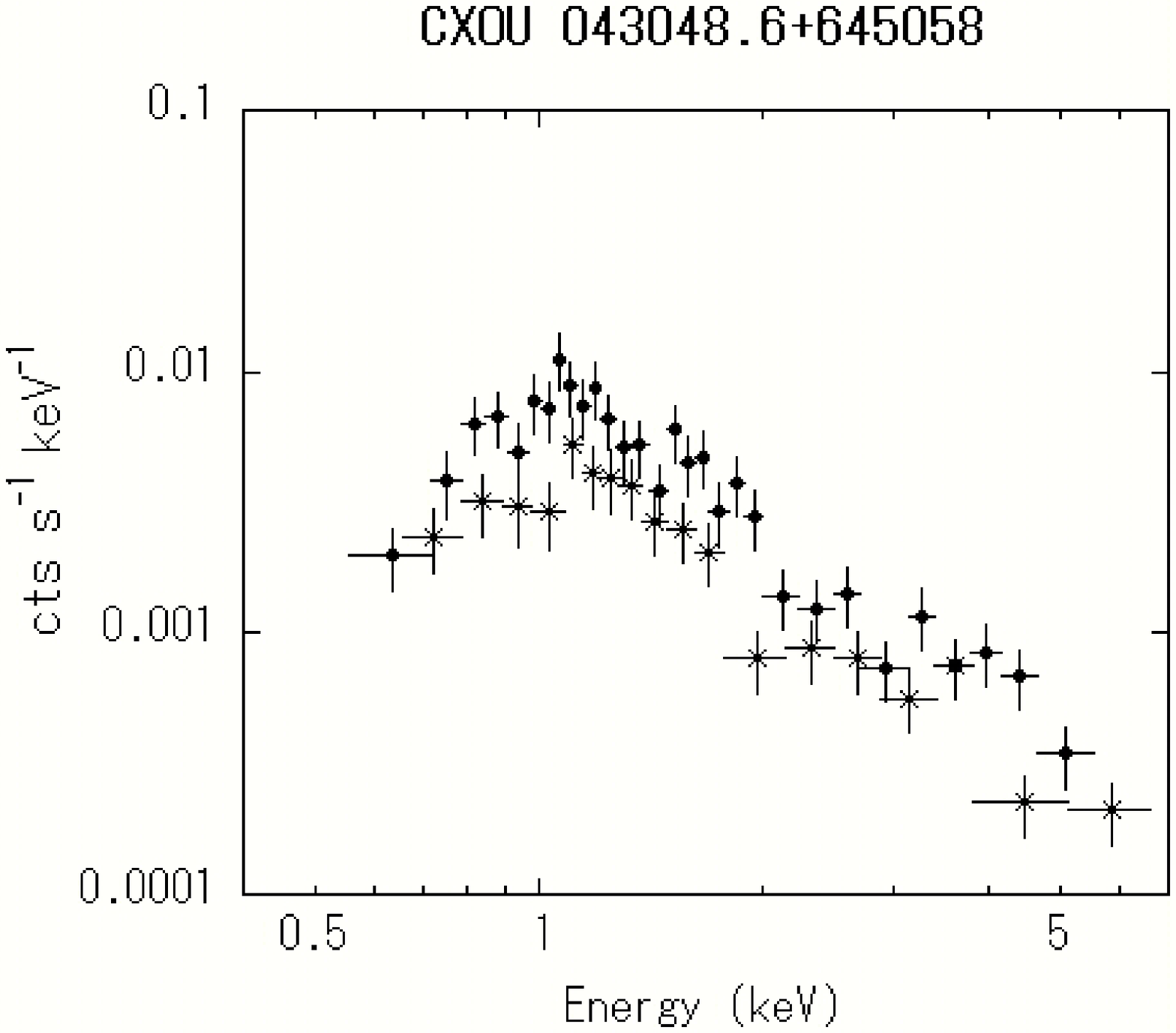}{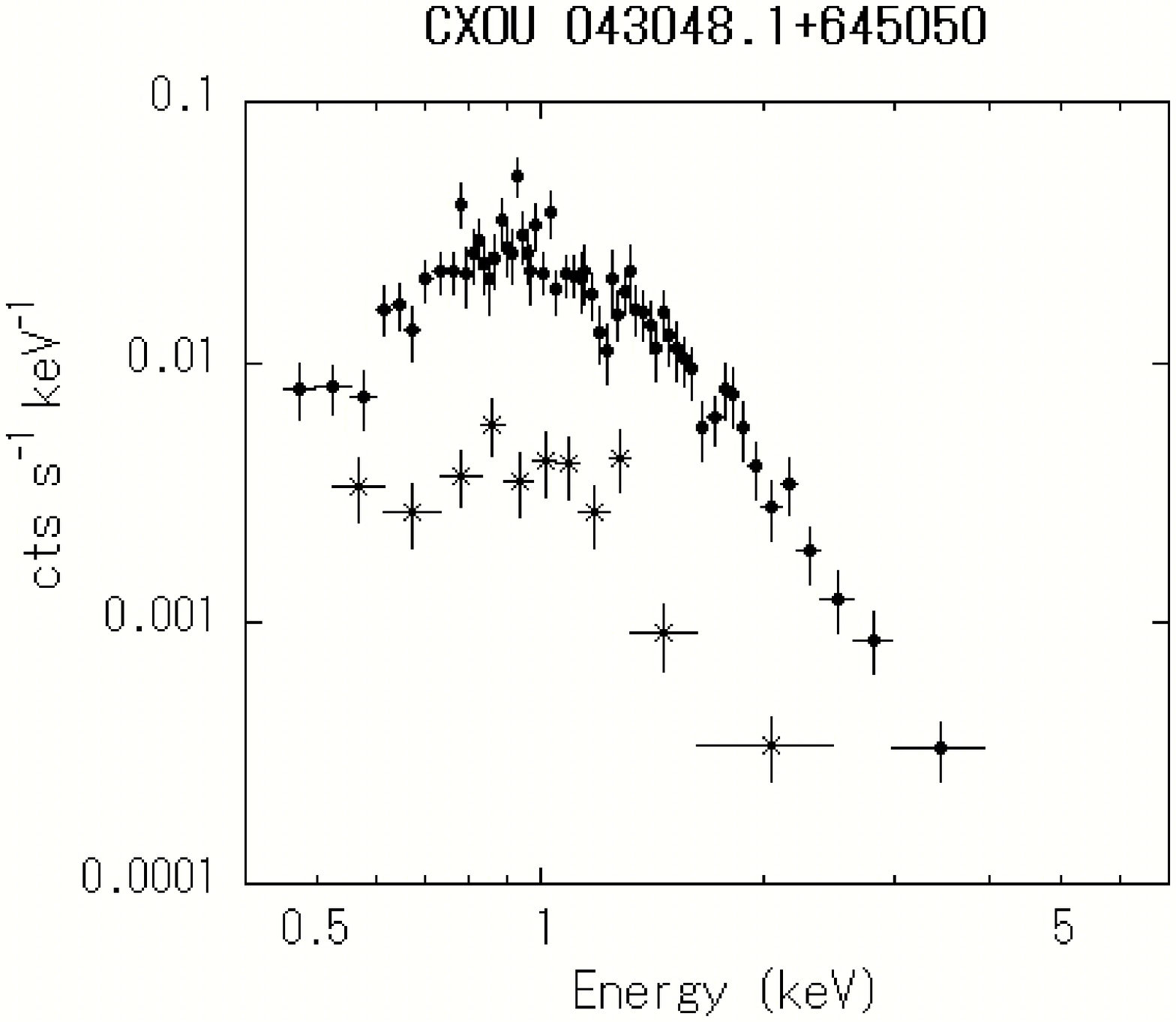}
\end{center}
\caption{The spectra of the bright X-ray sources in NGC 1569.
The X-ray spectra of the first half and the latter half of the observation
are presented by crosses and filled circles, respectively.
\label{fig3}
}
\end{figure}

\begin{figure}
\begin{center}
\plotone{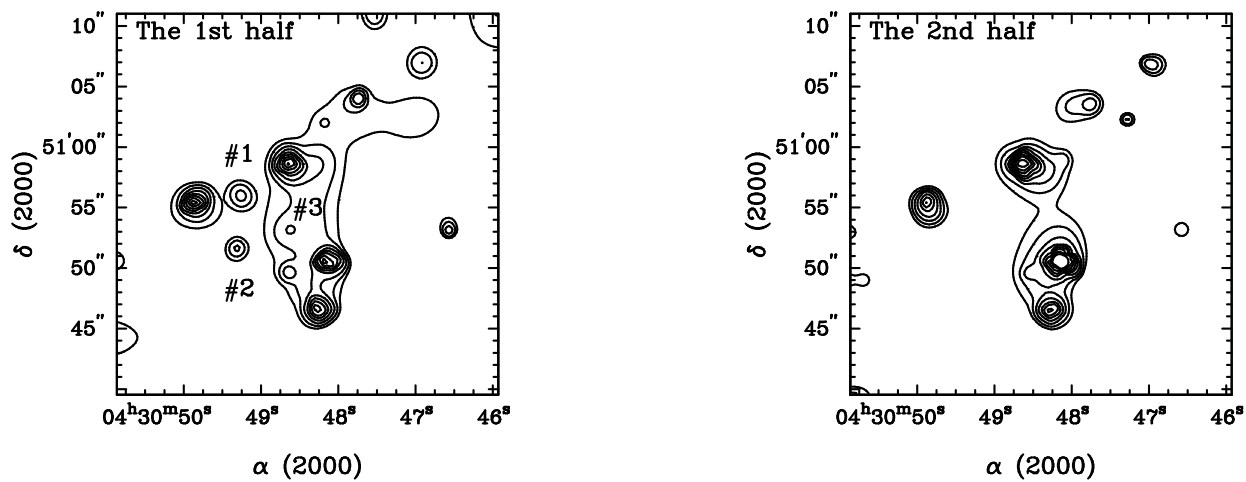}
\end{center}
\caption{X-ray images of the first half and the latter half of the observation.
The X-ray intensities of the faint sources labeled \#1, \#2 and \#3
have decreased in the latter half.
\label{fig4}
}
\end{figure}

\clearpage






\clearpage
\begin{deluxetable}{cccccccccc}
\tabletypesize{\scriptsize}
\tablecaption{The best fit parameters, fluxes and luminosities \label{tbl-1}}
\tablewidth{0pt}
\tablehead{
\colhead{source} & \colhead{OTZ}  & \colhead{net count rate}  & \colhead{$N_{\rm H}$} & \colhead{$\Gamma$}  &
\colhead{$\chi_{\nu}{}^{2}$(d.o.f.)} & \colhead{$F_{\rm X}^{\rm 0.5-10}$} &
\colhead{$L_{\rm X}^{\rm 0.5-10}$}
\\
\colhead{(1)} & \colhead{(2)}   & \colhead{(3)}   & \colhead{(4)} & \colhead{(5)} &
\colhead{(6)} & \colhead{(7)}   & \colhead{(8)}  \\
}
\startdata
CXOU 043048.1+645050 & 1st &  3.4$\pm$ 0.3  & 6.8$^{+4.7}_{-2.4}$  &5.95$^{+2.85}_{-0.73}$  & 1.3  (8)  &  1.0 &
0.6      \\
                     & 2nd & 25.2$\pm$ 0.7 & 5.5$^{+0.7}_{-0.6}$  &4.10$^{+0.35}_{-0.31}$ & 1.4 (56) &   8.9 &
5.3\    \\
CXOU 043048.6+645058 & 1st & 5.5$\pm$0.3  & 2.1 (fixed)  & 1.60$\pm$0.22  & 1.3 (16) &  5.4 &
3.1     \\
                     & 2nd & 9.9$\pm$ 0.5  &2.1 (fixed)  &1.51$^{+0.13}_{-0.14}$  & 1.2 (28) &  10.4 &
6.0       \\
\enddata






\tablecomments{Col.(1):source ID given in figure 1. 
Col.(2):observation time zone. ``1st'' is the first-half and ``2nd'' is the latter-half of this observation. 
Col.(3):net count rates of these sources in units of 10$^{-3}$ c s$^{-1}$.
Col.(4):absorbing column density in units of $10^{21}$cm$^{-2}$. 
Col.(5):photon index of power-law model. 
Col.(6):reduced chi-square.
Col.(7):observed fluxes from 0.5 to 10keV in units of 10$^{-14}$ergs s$^{-1}$cm$^{-2}$.
Col.(8):absorbed luminosities from 0.5 to 10keV in units of 10$^{37}$ergs s$^{-1}$.  
Errors are given for 90\% confidence.}
\end{deluxetable}



\clearpage
\begin{deluxetable}{cccccccccc}
\tabletypesize{\scriptsize}
\tablecaption{The best fit parameters of disk blackbody model \label{tbl-2}}
\tablewidth{0pt}
\tablehead{
\colhead{source} & \colhead{OTZ}  & \colhead{$N_{\rm H}$} & \colhead{$T_{\mathrm{in}}$}  &
\colhead{$\chi_{\nu}{}^{2}$(d.o.f.)} & \colhead{$L_{\rm bol}$}
\\
\colhead{(1)} & \colhead{(2)}   & \colhead{(3)}   & \colhead{(4)} & \colhead{(5)} & \colhead{(6)} \\
}
\startdata
CXOU 043048.1+645050 & 1st & 2.4$^{+2.5}_{-1.8}$ & 0.27$^{+0.13}_{-0.08}$ &
1.5 (8) & 1.3$^{+4.8}_{-0.7}$  \\
                     & 2nd & 2.1$\pm0.4$  &0.43$\pm$0.05 & 
1.5 (56)  & 5.5$^{+1.3}_{-0.6}$  \\

\enddata






\tablecomments{
Col.(1):source ID given in figure 1. 
Col.(2):observation time zone. ``1st'' is the first-half
and ``2nd'' is the latter-half of this observation. 
Col.(3):absorbing column density in units of $10^{21}$cm$^{-2}$. 
Col.(4):the temperature at the innermost disk boundary in units of keV. 
Col.(5):reduced chi-square. 
Col.(6):bolometric luminosity in units of 10$^{37}$ ergs s$^{-1}$ at 
the inclination angle of 0 deg. Errors are given for 90\% confidence.
}
\end{deluxetable}




\end{document}